\documentclass[12pt]{article}

\newif\ifnew
\newtrue

\usepackage{cite}
\usepackage[pdftex]{graphicx}
\usepackage{amsmath}
\usepackage{algorithmic}
\usepackage{array}
\usepackage{caption} 
\usepackage{amsmath}
\newtheorem{theorem}{Theorem}

\begin{document}

\title{     Reduction of the secret key length in the perfect cipher by data compression and randomisation}
\author{ Boris Ryabko  \\ \\{\small Federal Research Center for Information and Computational Technologies } \\ {\small Novosibirsk State  University} \\ }
\date{}
\maketitle

\thispagestyle{empty}

\begin{abstract}

Perfect ciphers have been a very attractive cryptographic tool ever since C. Shannon described them. Note that, by definition, if a perfect cipher is used, no one can get any information about the encrypted message without knowing the secret key. We consider the problem of reducing the key length of perfect ciphers, because in many applications the length of the secret key is a crucial parameter. This paper describes a simple method of key length reduction. This method gives a perfect cipher and is based on the use of data compression and randomisation, and the average key length can be made close to Shannon entropy (which is the key length limit). It should be noted that the method can effectively use readily available  data compressors (archivers).

\end{abstract}


\section{Introduction }

Perfect ciphers are very attractive to cryptography, and they have found many applications since C.~Shannon published his famous work \cite {sh} where he described such ciphers and proved that the so-called one-time pad (or Vernam cipher) is perfect.  The concept refers to secret-key cryptography involving three participants Alice, Bob and Eve, where Alice wants to send a message to Bob in secret from Eve, who has the ability to read all correspondence between Alice and Bob. To do this, Alice and Bob use a cipher with a secret key $k$ (i.e. a word from some alphabet), which is known to them in advance (but not to Eve).  When Alice wants to send some message $m$, she first encrypts $m$ using key $k$ and sends it to Bob, who in turn decrypts the received encrypted message using the key $k$.   Eve also receives the encrypted message and tries to decrypt it without knowing the key. The system is called  perfect,  if Eve,  with computers and other equipment of unlimited power and unlimited time, cannot obtain any information about the encrypted message.   
Not only did C.~Shannon provide a formal definition of perfect (or unconditional) secrecy, but he also showed that the so-called one-time pad (or Vernam cipher) is such a system. One of the specific properties of this system is the equivalence of the length of the secret key and the message (or its entropy). 
Quite often this property has limited practical application as many modern telecommunication systems forward and store megabytes of information.

The natural idea is to compress messages before encryption by lossless data compression to reduce the message length and hence the secret key, and then apply  a one-time pad. 
Note that, on the one hand, data compression before encryption is ``folk wisdom'' in cryptography, but on the other hand,  naïve approaches to joint compression and encryption have led to well-documented security breaches in real-world systems \cite{2}.

To illustrate the non-triviality of this problem, consider a toy example. Suppose Alice wants to send Bob a message from the set $M$,  wanting to use a one-time pad. Suppose Alice wants to save the bits of the secret key and decides to first encode the message using a data compressor and then encrypt the compressed sequence using the one-time pad and send it to Bob. Generally speaking, she will save the secret key, but Eve gets extra information (the length of the compressed message). Indeed, she knows the data compressor being used and therefore gets some information about the original message, especially for small $M$.

In this paper we propose a perfect cipher that uses data compression and randomisation. 
Schematically, the proposed cipher can be divided into three steps.  Denote all possible messages by $M$ and describe the first step as follows: 
  the original message $m \in M$ is compressed by some data compressor $\phi$ (let the resulting word 
$\phi(m) =x$
$ = x_1 … x_s$ and %
$ l = \max_{m \in M}  | \phi(m)| $). 
Second, the resulting word $x$ is encrypted with a one-time pad $\psi$ using a secret key of length $s$ bits (let   $\psi(x) =  y= y_1 y_2 ... y_s$ and, third, word $y$ is expanded into $l-s$ random bits, which are independent and obey the Bernoulli distribution with $p(0)=p(1)=1/2$. (So, the resulting word  is
$ y_1 y_2 ... y_s r_{s+1} ... r_l$, where $r_i$ are the random bits).  

Thus, the goal of this cipher is to reduce the length of the secret key used and hence it makes sense to use this cipher if the value of the secret bits is much greater than the value of the random bits.  It is worth noting that the proposed cipher is simple and, in fact, its complexity is determined by the data compression algorithm and/or the random bit generator.  In addition, any practically usable archiver (e.g. ZIP) can be used in conjunction with this cipher.
The average length of the secret key can be made close to the limit (i.e. Shannon entropy), in case the probability distribution of the encrypted messages is known or unknown. In the former case, the well-known Huffman code can be applied, in the latter, a universal code (or standard archiver).


\section{Data compression}
\subsection{Prefix-free and trimmed codes}

Let $A$ be a finite alphabet and $A^n$ be the sets of $n$-letter words  ($n \ge 1$) and $A^* = \cup_{n=1}^\infty A^n$.  For some $n \ge 1$, a code is a map $\phi: A^n \to \{0,1\}^*$. A  code $\phi$ is lossless if there exists $\phi^{-1}$ such that 
$\phi^{-1}(\phi(m)) = m$ for any $m \in A^n$.   
We will consider so-called prefix-free codes. It means that the set of all codewords  
$\{ \phi(m):  m \in A^n\}$ is prefix-free. 
(Recall that, by definition,  some set of words $U$ is  prefix-free if for all  words $u,v \in U$ neither $u$ is a prefix of $v$ nor $v$ is a prefix of $u$.)

Trimmed  codes are designed to convert any code that has several very long code words to shorten them so that the prefix-free code is translated into a prefix-free new code. More precisely,
let   $\lambda$ be a  code for elements from $M = m_1, ..., m_L ,  M \subset A^*$.  
For some code and $m \in M$,  the codeword length $|\lambda(m)| $ can be about $L$. 
The average length of the secrete key is determined  by the codeword lengths as well as  complexity of the cipher proposed  and    depends on the lengths of the codewords. Thus, there are situations where it is convenient to use codes for which the codeword length of all letters does not exceed $\lceil\log L\rceil+1$. (instead of about $L$).
 We call such codes  trimmed, and define one specific code with this property as follows: 
 if $\lambda$ is a code then  
\begin {equation}\label{tr}
\lambda^{tr}(a_i) =
  \begin{cases}
    0\, \lambda(a_i)       & \quad \text{if } |\lambda(a_i) |\le \lceil \log L \rceil \\
   1\,  bin_{\lceil \log L \rceil}(i) & \quad \text{if }   |\lambda(a_i) | > \lceil \log L \rceil  \, ,
  \end{cases}
\end{equation}
where $ bin_{\lceil \log L \rceil}(i)$ is a binary presentation of $i$ whose length is $\lceil \log L \rceil$. (For example, $bin_4(3) = 0011$).   We see that the maximal codeword length is not greater than $ \lceil \log L \rceil + 1 $.  (Also, note that there are  prefix-free codes for which the maximal codeword length is   $\lceil \log L \rceil $.)

Let us explain how to decode $\lambda^{tr}$.  First,  the decoder reads the first binary letter. If it is $0$, the decoder uses the codeword of the code $\lambda$ in order to find the encoded letter. If the first letter is $1$, the next $\lceil \log L \rceil$ letters contain the binary decomposition 
 of $i$, i.e. the letter is $m_i$.

\subsection{Data compression and the Shannon entropy} 
Ever since C. Shannon published his famous paper \cite{sh1}, it has been known that Shannon entropy is a lower bound on the average length of any lossless prefix-free code. 
In the proposed cipher, the average length of the secret key used is equal to the average length of the code word used to compress the data before encryption.  Therefore, let us briefly consider data compression methods separately for the cases of known and unknown statistics.

Suppose there is a set $A^n$, $n \ge 1$, and a probability distribution $p$ on $A^n$. The famous Huffman code has a minimum average codeword length lying in the interval $[h(p), h(p)+1 )$, where $h(p) = -
\sum_{v \in A^n}  p(v) \log_2 p(v) $ is Shannon entropy \cite{co}.   For some distribution $p$ the codeword length can be $|A^n| - 1$. There may be a situation where this is inconvenient and then a truncated code can be used. In this case, the average codeword length can be bounded from above by $h(p)+2$.

 Let us consider the data compression methods (or universal codes) for the case of unknown statistics.
Note that  nowadays there are many universal codes which are based on different ideas and approaches, among which we note the PPM universal code \cite{cleary1984data},  the arithmetic code \cite{rissanen1979arithmetic}), the Lempel-Ziv (LZ)  codes
\cite{ziv1977universal}, the Burrows-Wheeler transform \cite{burrows1994block} which is used along with the book-stack (or MTF) code \cite{ryabko1980data,bentley1986locally,ryabko1987technical},  
the  class of grammar-based codes \cite{kieffer2000grammar,yang2000efficient}
and some others \cite{drmota2010tunstall,LS,BRyabko:84, re3}. 
These codes are universal. This means that, asymptotically, the length of the compressed file goes to the  smallest possible value,  i.e.~the  Shannon entropy  ($h(\nu)$) per letter.  
It is worth noting that modern archivers are based on these codes and are also universal. (As far as a real computer program can satisfy the asymptotic properties.)

When some real data compressor $\phi$ 
is applied to long messages (say $m \in A^n$ and $n \ge 1000$), it may be difficult to find the value 
\begin {equation}\label{i}
l = \max_{u \in A^n} |\phi(u) |
\end{equation}
that is used for encryption.
In this case, a truncated code $\phi^{tr}$ can be used, since for this code the maximum codeword length is
 $ \lceil n \log |A| \rceil +1 $ and is known in advance, see (\ref{tr}).

\section{Description of the cipher}.
Let Alice want to send the word $m= m_1 ... m_n \in A^n$ to Bob and apply a cipher to hide $m$ from Eve.  Alice and Bob are going to use  the prefix-free data compressor $\phi$ and some bits of srcret key $k$ whose length is not less than $l$ in (\ref{i}).

 Alice finds $\phi(m)$ (let it be $x_1 ... x_s$) and calculates $y_1 = x_1 \oplus k_1, y_2= x_2 \oplus k_2 ... , y_s=x_s \oplus k_s$. Alice then generates $l-s$ of random independent binary digits $ r_1 r_2 .... r_{l-s}$ and constructs a encrypted word
\begin {equation}\label{sw}
E( m,k,r)           = y_1 ... y_s r_1 r_2 .... r_{l-s} .
\end {equation}
Alice then sends Bob the word $E( m,k,r)$.
While decoding it, Bob calculates $x_1 = y_1 \oplus k_1, x_2= y_2 \oplus k_2 ... $ and stops this process as soon as it gets a word of $v, v \in \{\phi(m), m \in A^n\}$. 
(Note that such a word $v$ is a  single word, since $\phi$ is prefix-free.) 
We denote this cipher by $C\&R$ because it is based on compression and randomisation.

The following statement describes the properies of the cipher $C\&R$.
\begin{theorem}\label{t1}.  
Let the cipher  $C\&R$ is applied to messages from some $M $ and $E$ be a set of all possible encrypted words (\ref{sw}). 
Then,  $C\&R$ is a perfect cipher, that is 
$$
P(m|e) = P(m)
$$
for all $m \in M, e \in E$ (where $P(m)$ and $P(m|e)$ are probability and conditional probability, correspondingly and the equality is as in the definition 
of a  perfect cipher \cite{sh1}).
\end{theorem}
{\it Proof of Theorem. } 
Let $\phi$ be a data compressor used in the considered cipher  $C\&R$ and 
let $m  $ and $e $ be any  elements from $ M$ and $  E$,   $ \phi(m) =     x_1x_2 ... x_s  $,   where $s= |\phi(m)|$. 

C.~Shannon showed that the cipher is perfect if and only if 
$$
P(e|m) = P(e)
$$
for all $m \in M, e \in E$ (where $P(m)$ and $P(m|e)$, see \cite{sh1}).
We will use this theorem and first estimate $P(e|m)$ as follows: 
$$
P(e|m) =  \prod_{i=1}^{s} P\{k_i = x_i  \oplus e_i \} \,  (\prod_{j= s+1}^l   P\{ r_j = e_j \} ) \,\, , 
$$
where $l$ is defined in (\ref{i}).  Indeed, this equation shows that $x_i \oplus k_i = x_i$ $\oplus ( x_i \oplus e_i) = e_i$, and the random variables $x_i$ and $r_j$ obey the Bernoulli distribution, hence $P( e|m) = 2^{-l}$.
Now we estimate $P(e)$. 
$$P(e) =
\sum_{m\in M}  P(m) ( \prod_{i=1}^{|\phi(m)|} P\{k_i = x_i  \oplus e_i \} \,  (\prod_{j= |\phi(m)|+1}^l   P\{ r_j = e_j \} ) ) \, ,
$$
 where $\phi(m) = x_1 ... x_{|\phi(m)|} $. Repeating the estimates from the previous case, we see that
$ P(e) = \sum_{m\in M}  P(m) \, 2^{-l}$.  Hence,  $ P(e) = 1 \,\,  2^{-l} = 2^{-l}$. So,  $ P(e) = P(e|m)$ and the theorem is proven.


\end{document}